\documentclass{ws-mpla}

\begin{document}

\markboth{J. Ben Geloun, J. Govaerts and M. N. Hounkonnou}
{A $(p,q)$-deformed Landau problem}

\catchline{}{}{}{}{}

\title{A $(p,q)$-DEFORMED LANDAU PROBLEM\\
IN A SPHERICAL HARMONIC WELL:\\
SPECTRUM AND NONCOMMUTING COORDINATES}

\author{\footnotesize JOSEPH BEN GELOUN$^1$,
JAN GOVAERTS$^{1,2,}$\footnote{On sabbatical
leave from the Center for Particle Physics and Phenomenology (CP3),
Institute of Nuclear Physics, Universit\'e catholique de Louvain,
2, Chemin du Cyclotron, B-1348 Louvain-la-Neuve, Belgium.}
\ AND M. NORBERT HOUNKONNOU$^{1}$}

\address{$^{1}$International Chair in Mathematical Physics
and Applications (ICMPA--UNESCO)\\
072 B.P. 50  Cotonou, Republic of Benin\\
jobengeloun@yahoo.fr,\ \ norbert$_-$hounkonnou@cipma.net}

\address{$^{2}$Department of Theoretical Physics, School of Physics\\
The University of New South Wales, Sydney NSW 2052, Australia \\
Jan.Govaerts@fynu.ucl.ac.be}

\maketitle

\pub{Received (Day Month Year)}{Revised (Day Month Year)}

\begin{abstract}
A $(p,q)$-deformation of the Landau problem in a spherically symmetric
harmonic potential is considered. The quantum spectrum as well as
space noncommutativity are established, whether for
the full Landau problem or its quantum Hall projections.
The well known noncommutative geometry in each Landau level is recovered
in the appropriate limit $p,q=1$. However, a novel noncommutative algebra
for space coordinates is obtained in the $(p,q)$-deformed case, which could also
be of interest to collective phenomena in condensed matter systems.

\end{abstract}

\ccode{PACS Nos.: 02.40.Gh, 02.20.-a, 02.20.Uw}

\section{Introduction}	

Noncommuting spatial coordinates and fields can (approximately) be realized
in actual physical situations. Landau models and their quantum Hall limit
have become the focus of intense research activity as a physical
realization of the simplest example of noncommutative geometry.\cite{riccardi}$^{-}$\cite{cappeli}
Similar structures also arise in specific approaches towards a theory of
quantum gravity, such as M-theory in the presence of background fields\cite{SW}
or tentative formulations of relativistic quantum theories of gravity
through spacetime noncommutativity.\cite{conf}
 
As is well known,\cite{jackiw} given a point particle 
of mass $m$, charge $\bar{q}$ and position $\vec{r}=(x,y)$
moving in a plane in the presence of a constant external
magnetic field $B$ perpendicular to that plane,
the spectrum of the quantized theory is organized into infinitely
degenerate Landau levels, with separation ${\cal O}(\bar{q}B/m)$.
The limit $B\to \infty$ effectively projects onto
the lowest Landau level and is equivalent to a negligibly small mass
$m$, {\it i.e.} $m\to 0$. Consequently in each of the projected Landau
levels, one obtains a noncommuting algebra for the space coordinates,
\begin{equation}
\label{noc}
\left[x,y\right]=-\frac{i\hbar}{\bar{q}B}.
\end{equation}
Historically, this is the Peierls substitution rule 
introduced seventy years ago.\cite{peierls} 
As a matter of fact, this noncommuting character arises already
at the classical level in terms of the Dirac brackets associated to the second-class
constraints that follow upon taking the limit $m\to 0$ in the classical
Hamiltonian formulation of the dynamics of the Landau problem and requiring
finite energy configurations.

A deformation of the Quantum Hall Effect has been already considered
within the Moyal approach and the associated star product for space(time) noncommutativity,
the quantum algebra and group formalisms\cite{riccardi,suss}$^-$\cite{cappeli} then providing useful 
frameworks for developments along such lines. More general deformations and their
ensuing space noncommutativity structures should also be of physical
interest and provide further potential applications. As was discussed 
in Ref.~\refcite{scholtz}, it appears possible to associate to a fully interacting
system in presence of a magnetic field $B$ new types of
noncommutative geometries when restricting the space of quantum states
to its lower energy sector. From that point of view, when some complex dynamics may, 
in some regime or within some physical limits, be approximated by a model
which itself is exactly solvable, explicit and analytic evaluations 
performed within the latter are so much more efficient and often
more transparent and instructive than any approximate or perturbative 
solution or numerical simulation of the original complex dynamics.

In this Letter, we wish to point out that in the same spirit,
and based still on the Landau problem, other types of deformations
of its quantum algebra also provide for yet other realizations of
noncommuting space coordinates, which encompass in specific limits
the above by now standard scheme (\ref{noc}) in which the space coordinate
commutator is purely a constant factor multiplying the unit operator.

Given the two parameter $(p,q)$-deformation of the
harmonic oscillator or Fock space algebra,\cite{chak,vin}
we investigate the space noncommutativity which is implied
in a $(p,q)$-deformed Landau and quantum Hall system.
One of the aims of this contribution is to show that besides the type
of noncommutative space geometry associated to the Landau problem and its
quantum Hall effect which appears for instance in string theory
in specific limits in the presence of background fields, 
consistent and potentially interesting alternatives also exist 
and deserve to be considered as well with equal interest. 

\section{The $(p,q)$-Deformed Extended Landau Problem}

Consider the Lagrange function
\begin{eqnarray}
\label{lag}
L= {\textstyle\frac{1}{2}}m\dot{\vec{r}}^{\;2}
 + \bar{q} \vec{E}\cdot \vec{r} + \bar{q} \dot{\vec{r}}\cdot 
 \vec{A}(\vec{r})
 - {\textstyle\frac{1}{2}}k \vec{r}^{\;2},
\end{eqnarray}
describing the nonrelativistic motion of a charged
massive particle confined in the $(x,y)$-plane,
submitted to static background fields defined by
a planar constant electric field $\vec{E}=(E_{1},E_{2},0)$ and the
vector potential $\vec{A}(\vec{r})=\vec{B}\times \vec{r}/2$
associated to a constant magnetic field $\vec{B}=(0,0,B)$
perpendicular to the $(x,y)$-plane, considered in the symmetric gauge.
A spherically symmetric interacting harmonic potential is also included,
with angular frequency $\omega$ and stiffness constant $k=m\omega^{2}$.

Introducing the displaced coordinates defined by
$\vec{R}(t)= \vec{r}(t)- \bar{q}\vec{E}/k$,
which minimize the total potential energy, 
the Hamiltonian associated with (\ref{lag}) can be put in the
form 
\begin{eqnarray}
\label{ham}
H&=&{\textstyle\frac{1}{2m}}\left[\vec{P} - 
\bar{q}\vec{A}\left(\vec{R}+{\textstyle\frac{\bar{q}}{k}}\vec{E}\right)\right]^{\;2}
 + {\textstyle\frac{1}{2}}k \vec{R}^{\;2} - {\textstyle\frac{\bar{q}^2}{2k}} \vec{E}^{\;2},
 \end{eqnarray}
$\vec{P}=(P_X,P_Y,0)$ being the momenta conjugate to the displaced coordinates
$\vec{R}=(X,Y,0)$. Equivalently, the Hamiltonian (\ref{ham}) expands as
\begin{eqnarray}\label{ham2}
H&=& {\textstyle\frac{1}{2m}}(\pi_{X}^{2}+\pi_{Y}^{2})
+{\textstyle\frac{\bar{q}B}{2m}}(\pi_{X}Y-\pi_{Y}X)
+ {\textstyle\frac{1}{2}}m \Omega^{2} (X^{2}+ Y^{2})
- {\textstyle\frac{\bar{q}^2}{2k}} \vec{E}^{\;2},
\end{eqnarray}
where $\vec{\pi}=(\pi_{X},\pi_{Y},0)=
 \vec{P}+ [\bar{q}^{2}/(2k)] \vec{E}\times \vec{B}$
and  $\Omega=[\omega^{2}+ (\bar{q}B)^{2}/(4m^{2})]^{1/2}$.
Note that the phase space coordinates $(\vec{R},\vec{\pi})$ are also
canonically conjugate, with as Poisson brackets the canonical ones,
$\{ X,P_X\}=1=\{ Y,P_Y\}$ or $\{ X,\pi_X\}=1=\{ Y,\pi_Y\}$.
Introducing, for $Z=X,Y$,\footnote{For convenience, the $\hbar$ factor relevant
to the quantized system is already introduced in these expressions simply as a dimensional
numerical constant, even though they still refer to classical quantities.}
\begin{eqnarray}
a_{Z}=\left[{\textstyle\frac{m\Omega}{2\hbar}}\right]^{1/2}
[Z + {\textstyle\frac{i}{m\Omega}}\pi_{Z}],\;\;\;\;
a_{\pm}={\textstyle\frac{1}{\sqrt{2}}}[a_{X} \mp i a_{Y}],
\end{eqnarray}
and using the above Poisson brackets, the only nonvanishing
brackets for the modes $a_{\pm}$ are
\begin{eqnarray}\label{aa}
\{a_{\pm}, a_{\pm}^{\ast}\}=-{\textstyle\frac{i}{\hbar}},
\end{eqnarray}
$a^{\ast}_\pm$ being the complex conjugate of the variable $a_\pm$.
Inverting these relations, and after some algebra, one finds from (\ref{ham2})
\begin{eqnarray}\label{hamo}
H\!&=&{\textstyle\frac{\hbar\Omega}{2}}
[a_{+}a_{+}^{\ast} + a_{+}^{\ast}a_{+}
+ a_{-}a_{-}^{\ast}+ a_{-}^{\ast}a_{-}] \cr
&-&
{\textstyle\frac{\hbar\bar{q}B}{4m}}
[a_{+}a_{+}^{\ast} + a_{+}^{\ast}a_{+}
 - (a_{-}a_{-}^{\ast} + a_{-}^{\ast}a_{-})]
 - {\textstyle\frac{[\bar{q}\vec{E}]^{\;2}}{2k}}.
\end{eqnarray}

Ordinary canonical quantization thus leads to two commuting Fock algebras,
$[a_\pm,a^\dagger_\pm]=\mathbb{I}$,
each associated to each of the two helicity sectors, which diagonalize
the Hamiltonian. In particular, in the absence of both the harmonic well potential,
$k=0$, and\footnote{Indeed the naive limit $k=0$ requires to first set
$\vec{E}=\vec{0}$.} the external electric field, $\vec{E}=\vec{0}$,
energy eigenvalues are infinitely degenerate on account of translation
invariance. Without loss of generality, let us assume henceforth that $\bar{q}B/2m>0$,
in which case\footnote{If $\bar{q}B/2m<0$, the role of the two helicity sectors is simply
exchanged.} the infinite degeneracy for each value of $n_-\ge 0$ is labelled by the angular momentum
$L=(X\pi_Y-Y\pi_X)$ with eigenvalues $\hbar\ell$ where $\ell=n_+-n_-\ge -n_-$, $n_+\ge 0$
(resp. $n_-\ge 0$) counting the excitation level in the $a_+^\dagger$
(resp. $a_-^\dagger$) modes, while energy eigenvalues are
$E(n_+,n_-)=\hbar(\bar{q}B/m)(n_-+1/2)$.

In this Letter however, quantization of the above classical algebra
will proceed with the two parameter deformation of the Fock algebra
introduced by Chakrabarty and Jagannathan,\cite{chak}
namely through the so-called $(p,q)$-oscillator quantum algebra.
Consider two couples of parameters $(p_{\pm},q_{\pm})$,
with $p_\pm$ and $q_\pm$ both real such that $p_{\pm}>1$ and 
$|q_{\pm}|<1$, and the generators $a_{\pm}$, $a^{\dagger}_{\pm}$
and $N_{\pm}$ which obey\cite{chak,vin} 
\begin{eqnarray}
\label{pqpm}
&&[a_{\pm},a^{\dag}_{\pm}]_{q_{\pm}}:=a_{\pm}a_{\pm}^{\dag}
-q_{\pm}a_{\pm}^{\dag}a_{\pm}=p_{\pm}^{-N_{\pm}},\;\;
[a_{\pm},a_{\pm}^{\dag}]_{p_{\pm}}:=
a_{\pm}a_{\pm}^{\dag}
-{\textstyle\frac{1}{p_{\pm}}}
a^{\dag}_{\pm}a_{\pm}=q_{\pm}^{N_{\pm}},\cr
&&[N_{\pm},a_{\pm}]=-a_{\pm},\;\;
[N_{\pm},a_{\pm}^{\dag}]=a_{\pm}^{\dag}.
\end{eqnarray}
Then, (\ref{pqpm}) defines two decoupled $(p,q)$-deformed 
oscillators.{\footnote{The limit $p_{\pm}\to 1^{+}$ yields the 
$q_{\pm}$-oscillator of Arik and
Coon;\cite{ar} likewise $p_{\pm}=q_{\pm}$ gives the $q_{\pm}$-deformed
oscillator algebra of Biedenharn and MacFarlane.\cite{far}
Finally, the algebras (\ref{pqpm}) reduces to ordinary
harmonic oscillator algebras as $q_{\pm}\to 1$ for 
$p_{\pm}=1$ or $p_{\pm}=q_{\pm}$.}}
The two mode Fock-Hilbert space of states
of these algebras is nothing but the usual tensor product
of two $(p,q)$-Fock-Hilbert spaces\cite{chak} 
each associated to each of the helicity sectors of the Landau problem
and given by the vacuum $|0\rangle =|0_{+},0_{-}\rangle$
annihilated by $a_{\pm}$ and normalized such that
$a_{\pm}|0\rangle=0,\;\langle 0|0\rangle=1,\;$ and the further
orthonormalized states defined by
\begin{eqnarray}
\label{pqreps}
|n_{+},n_{-}\rangle
= \left([n_{+}]_{p_{+},q_{+}}![n_{-}]_{p_{-},q_{-}}!\right)^{-1/2}
[a_{+}^{\dag}]^{n_{+}}[a_{-}^{\dag}]^{n_{-}}|0\rangle.
\end{eqnarray}
The symbols 
$[n]_{p_{\pm},q_{\pm}}=
\left(p_{\pm}^{-n}-q_{\pm}^{n}\right)/\left(p_{\pm}^{-1}-q_{\pm}\right)$
are called $(p_{\pm},q_{\pm})$-basic numbers with, by convention, 
$[0]_{p_{\pm},q_{\pm}}=0$,
and their $(p_{\pm},q_{\pm})$-factorial 
$[n]_{p_{\pm},q_{\pm}}!=[n]_{p_{\pm},q_{\pm}}[n-1]_{p_{\pm},q_{\pm}}!$, 
with $[0]_{p_{\pm},q_{\pm}}!=1$.
The actions of the operators $a_{\pm}$, $a_{\pm}^{\dag}$ and
$N_{\pm}$ on the total Hilbert space readily follow through the usual 
tensor product rules and the relations for one mode action
\begin{eqnarray}
&&
a_{\pm}|n_\pm,n_\mp\rangle=\left([n_\pm]_{p_{\pm},q_{\pm}}\right)^{1/2}|n_\pm-1,n_\mp\rangle,\; \cr
&&
a_{\pm}^{\dag} |n_\pm,n_\mp\rangle = \left([n_{\pm}+1]_{p_{\pm},q_{\pm}}\right)^{1/2}|n_\pm+1,n_\mp\rangle,
\cr&& N_{\pm}|n_\pm,n_\mp\rangle  =n_\pm|n_\pm,n_\mp\rangle.
\end{eqnarray}
There exist formal $(p_{\pm},q_{\pm})$-number operators
denoted by $[N_{\pm}]_{p_{\pm},q_{\pm}}$, or simply by $[N_{\pm}]$ 
when no confusion occurs,
and defined by $[N_{\pm}]=a_{\pm}^{\dag}a_{\pm}$. 
One has $[N_{\pm}]|n_+,n_-\rangle= [n_{\pm}]|n_+,n_-\rangle$.
A basic $q$-arithmetics shows that
$a_{\pm}a_{\pm}^{\dag}|n_+,n_-\rangle=[N_{\pm}+1]|n_+,n_-\rangle$
so that the operators $a_{\pm}a_{\pm}^{\dag}=q_{\pm}[N_{\pm}]+p_{\pm}^{-N_{\pm}}$
are simply denoted by $[N_{\pm}+1]$. Hence, (\ref{pqreps})
provides a well defined Fock-Hilbertian representation space of the
algebra (\ref{pqpm}).

The $(p,q)$-deformed quantum Hamiltonian associated to (\ref{hamo})
is given by
\begin{eqnarray}
\label{hpq}
H_{p,q}&=&\ \ {\textstyle\frac{1}{2}}\hbar\Omega
\sum_{\epsilon=\pm}
\left( p_{\epsilon}^{-N_{\epsilon}} + 
(1+q_{\epsilon})[N_{\epsilon}]_{p_\epsilon,q_\epsilon}\right)\\
&&-\ {\textstyle\frac{1}{2}}\hbar{\textstyle\frac{\bar{q}B}{2m}}
\sum_{\epsilon=\pm}\,\epsilon
\left(p_{\epsilon}^{-N_{\epsilon}} + (1+q_{\epsilon})[N_{\epsilon}]_{p_\epsilon,q_\epsilon}\right)\,-\,
{\textstyle\frac{\bar{q}^2}{2k}}\vec{E}^{\;2},
\end{eqnarray}
which is thus diagonalized on the basis $|n_+,n_-\rangle$ of the two mode Fock-Hilbert
space with eigenvalues
\begin{eqnarray}
E_{p,q}(n_{+},n_{-})&=&\ \  {\textstyle\frac{1}{2}}\hbar\Omega
\sum_{\epsilon=\pm}
\left( p_{\epsilon}^{-n_{\epsilon}} +
(1+q_{\epsilon})[n_{\epsilon}]_{p_\epsilon,q_\epsilon}\right)\cr
&& -\ {\textstyle\frac{1}{2}}\hbar{\textstyle\frac{\bar{q}B}{2m}}
\sum_{\epsilon=\pm}\,\epsilon
\left(p_{\epsilon}^{-n_{\epsilon}} + 
(1+q_{\epsilon})[n_{\epsilon}]_{p_\epsilon,q_\epsilon}\right)\,-\,
{\textstyle\frac{\bar{q}^2}{2k}}\vec{E}^{\;2} .
\label{eq:spectrum}
\end{eqnarray}
Note that in the absence of the electric field $\vec{E}$ and the harmonic well, 
the $(p,q)$-deformed
Landau problem remains infinitely degenerate in the angular momentum 
$\ell=n_+-n_-\ge -n_-$
for each of the Landau levels distinguished by $n_-\ge 0$. 
Indeed, it may appear to be a matter of choice
whether the Hamiltonian and angular momentum operators in the 
$(p,q)$-deformed case are defined in terms
of the level operators $N_\pm$ or the combinations
$a^\dagger_\pm a_\pm + a_\pm a^\dagger_\pm=[N_\pm]_{p_\pm,q_\pm}+[N_\pm+1]_{p_\pm,q_\pm}$.
However, the above choice seems to be most natural in order to retain, on the one hand, 
a degenerate Landau problem when both $k=0$ and $\vec{E}=\vec{0}$, and on the other hand, 
an angular momentum spectrum which is integer valued (modulo the $\hbar$ factor)
as required for single-valuedness of state wave functions.
Thus the $(p,q)$-deformed angular momentum operator is defined by
$L_{p,q}=\hbar\left(N_+-N_-\right)$.
As it should, $H_{p,q}$ and $L_{p,q}$ indeed define commuting operators to be diagonalized simultaneously.
This is also the necessary choice for $L_{p,q}$ to belong to the 
$(p,q)$-deformed SU$_{p,q}$(2) algebra
built from the above two $(p,q)$-deformed Fock algebras in a manner analogous to that which leads to
a SU(2) algebra associated to two Fock algebras through the well known Jordan-Wigner construction.\cite{chak}

\section{Space Noncommutativity}

Given the above choice for a $(p,q)$-deformed quantum realization of the classical Landau problem,
it is to be expected that even before projecting onto any given Landau level labelled by a
specific value for $n_-\ge 0$, the quantum cartesian plane coordinates
\begin{eqnarray}
X={\textstyle\frac{1}{2}}
[{\textstyle\frac{\hbar}{m\Omega}}]^{1/2}
[a_{+}+a_{-}+a_{+}^{\dag}+a_{-}^{\dag}],\;\;\; 
Y={\textstyle\frac{i}{2}}
[{\textstyle\frac{\hbar}{m\Omega}}]^{1/2}
[a_{+}-a_{-}-a_{+}^{\dag}+a_{-}^{\dag}],
\end{eqnarray}
are no longer commuting operators whether simply for a $q$-deformation
or more generally given any real values for the pairs
$(p_\pm,q_\pm)$ such that $p_\pm>1$ and $|q_\pm|<1$.
A direct evaluation of their commutator readily yields
\begin{eqnarray}
\label{xy}
[X,Y]={\textstyle\frac{i\hbar}{2m\Omega}}
\left\{((1-q_+)[N_+]_{p_+,q_+}-p^{-N_+}_+)-
((1-q_-)[N_-]_{p_-,q_-}-p^{-N_-}_-)\right\}
\end{eqnarray}
a quantity which indeed does not vanish, unless one has both $p_\pm=1$
and $q_\pm=1$. Likewise, even when considering alternative commutators
for the cartesian coordinates, such as
$[X,Y]_{q_\pm}=XY-q_\pm YX$ and $[X,Y]_{p_\pm}=XY-p^{-1}_\pm YX$,
a direct evaluation of these quantities shows that a similar conclusion
applies to these choices as well. Consequently, the $(p,q)$-deformed Landau
problem is associated to a noncommutative quantum geometry in the Euclidean
plane, characterized by the commutator (\ref{xy}) which itself is now a nontrivial
and nonconstant operator, in contradistinction to (\ref{noc}) which applies only
in the quantum Hall limit of the Landau level projected quantities.

It is also of interest to establish the commutator of the cartesian
coordinate operators projected onto any of the Landau levels, thereby
deriving the quantum geometry associated to the $(p,q)$-deformed 
quantum Hall effect. The projector onto the Landau level $n^0_-$ is
simply
\begin{eqnarray}
\mathbb{P}(n^0_-)= \sum_{n_+=0}^{\infty}|n_+,n^0_-\rangle \langle n_+,n^0_-|.
\end{eqnarray}
The projected quantum cartesian coordinates are thus, with $Z=X,Y$,
\begin{eqnarray}
\bar{Z}= \mathbb{P}(n^0_-)Z\,\mathbb{P}(n^0_-)=
\sum_{n_+,m_+=0}^{\infty}
|n_+,n^0_-\rangle \langle n_+,n^0_-|Z |m_+,n^0_-\rangle \langle m_+,n^0_-|.
\end{eqnarray}
A direct evaluation then finds\footnote{Likewise, one could also list
the results for $[\bar{X},\bar{Y}]_{q_\pm}$ or
$[\bar{X},\bar{Y}]_{p_\pm}$, with similar conclusions.}
\begin{equation}
\left[\bar{X},\bar{Y}\right]=-{\textstyle\frac{i}{2}}
{\textstyle\frac{\hbar}{m\Omega}}
\sum_{n_+=0}^\infty\,|n_+,n^0_-\rangle
\left\{[n_++1]_{p_+,q_+}-[n_+]_{p_+,q_+}\right\}\langle n_+,n^0_-|,
\label{pxp}
\end{equation}
or equivalently,
\begin{eqnarray}
\left[\bar{X},\bar{Y}\right]&=&
{\textstyle\frac{i\hbar}{2m\Omega}}\sum_{n_+=0}^\infty\,
[n_++1]_{p_+,q_+}
\left\{|n_++1,n^0_-\rangle \langle n_++1,n^0_-|-
|n_+,n^0_-\rangle \langle n_+,n^0_-|\right\} \cr
&=&{\textstyle\frac{i\hbar}{2m\Omega}}
\left((1-q_+)[N_+]_{p_+,q_+}-p^{-N_+}_+\right)\,\mathbb{P}(n^0_-).
\label{eq:pxp2}\end{eqnarray}
Again, this results fails to reduce to a constant,
proportional to the unit operator. However, in the double
limit such that both $p_+=1$ and $q_+=1$, but only in that case,
one recovers of course the result
characteristic of the quantum Hall effect in the plane, namely
\begin{equation}
\left[\bar{X},\bar{Y}\right]=-{\textstyle\frac{i\hbar}{2m\Omega}}\,
\mathbb{P}(n^0_-).
\end{equation}
In the specific instance that $\vec{E}=\vec{0}$ and $k=0$, 
one then has (\ref{noc}).

Hence $(p,q)$-deformations of the Fock algebra also lead to continuously
$(p,q)$-deformed quantum geometries in the context of the Landau
problem and its Landau level projected quantum Hall cousin,
parametrized by a set of four real parameters $(p_\pm,q_\pm)$ such that
$p_\pm>1$ and $|q_\pm|<1$. It is only in the double limit $p_\pm=1$
and $q_\pm=1$ that the quantum geometry of the $(p,q)$-deformed Landau
problem becomes commuting, whereas that of the quantum Hall cousin
remains noncommuting, however then with a constant commutator for
the cartesian coordinates in the plane.

\section{The $(p,q)$-Deformed Landau Problem as an Approximation Model for Interactions}

In fact, the result in (\ref{eq:pxp2}) is reminiscent of a similar
one in Ref.~\refcite{scholtz} and of the programme outlined in 
Refs.~\refcite{scholtz} and \refcite{scholtz2}
which suggests that interactions may be traded for noncommutative geometry in
certain regimes of interaction energies. Following the reasoning of 
Ref.~\refcite{scholtz},
let us consider a two dimensional system described by the Lagrange function
\begin{equation}
L_0={\textstyle\frac{1}{2}}m_0\dot{\vec{r}}^{\;2}-V(|\vec{r}\,|)+
\bar{q}_0\dot{\vec{r}}\cdot\vec{A}_0(\vec{r}\,),
\end{equation}
with $\vec{A}_0(\vec{r}\,)$ the vector potential associated to
a constant magnetic field $\vec{B}_0$ perpendicular to the plane,
expressed in the symmetric gauge, and $V(|\vec{r}\,|)$ some given
spherically symmetric interacting potential.

In the ordinary configuration space representation in radial coordinates, the
time independent Schr\"odinger equation reduces to
\begin{equation}
\left\{-\frac{\hbar^2}{2m_0}
\left[\frac{d^2}{dr^2}+\frac{1}{r}\frac{d}{dr}-\frac{\ell^2}{r^2}\right]
-\hbar\frac{\bar{q}_0 B_0}{2m_0}\ell
+\frac{\bar{q}^2_0 B^2_0}{8m_0}r^2+V(r)\right\}\,R_{n,\ell}(r)=
E_{n,\ell}\,R_{n,\ell}(r),
\label{eq:Schro}
\end{equation}
for any energy eigenstate $|n,\ell\rangle$ of definite angular momentum $\hbar\ell$
such that,
\begin{equation}
\langle \vec{r}\,|n,\ell\rangle=\psi_{n,\ell}(\vec{r}\,)=R_{n,\ell}(r)\,e^{i\ell\theta} .
\end{equation}
Given our implicit assumption that $\bar{q}_0 B_0/2m_0>0$, eigenstates
are such that $\ell\ge -n$ where $n=0,1,2,\cdots$ labels the solutions
to (\ref{eq:Schro}) of increasing energies $E_{n,\ell}$. These quantum numbers
are related to the helicity modes of the Landau problem through $\ell=n_+-n_-$
and $n=n_-$. As discussed in Ref.~\refcite{scholtz}, 
the Landau level projected cartesian coordinate operators
\begin{equation}
\bar{x}=\mathbb{P}(n_0)x\,\mathbb{P}(n_0),\quad
\bar{y}=\mathbb{P}(n_0)y\,\mathbb{P}(n_0),
\;\;{\rm where}\;\;
\mathbb{P}(n_0)=\sum_{\ell=-n_0}^\infty\,
|n_0,\ell\rangle \langle n_0,\ell|,
\end{equation}
possess the commutator
\begin{equation}
\left[\bar{x},\bar{y}\right]=2i\sum_{\ell=-n_0}^\infty
\left|\Omega_{\ell,\ell+1}(n_0)\right|^2\,
\left\{|n_0,\ell+1\rangle \langle n_0,\ell+1|\,-\,
|n_0,\ell\rangle \langle n_0,\ell|\right\},
\label{eq:pxp3}
\end{equation}
where
\begin{equation}
\Omega_{\ell,\ell+1}(n_0)=\pi\int_0^\infty dr\,r^2\,
R^*_{n_0,\ell}(r)\,R_{n_0,\ell+1}(r),
\end{equation}
it being understood that the energy eigenstates $|n,\ell\rangle$ have been
normalized, $2\pi\int_0^\infty dr\,r^2 R^*_{n,\ell}(r)\,R_{n,\ell}(r)= 1$.
Comparing (\ref{eq:pxp2}) and (\ref{eq:pxp3}) suggests the correspondence
\begin{equation}
\left|\Omega_{\ell,\ell+1}(n_0)\right|^2\ \longleftrightarrow\
\frac{\hbar}{4m\Omega}[\ell+n_0+1]_{p_+,q_+}=
\frac{\hbar}{4m\Omega}\frac{q^{\ell+n_0+1}_+
- p^{-(\ell+n_0+1)}_+}{q_+-p^{-1}_+}.
\end{equation}

More generally, in the same spirit as advocated in Refs.~\refcite{scholtz} and \refcite{scholtz2}
it appears possible that a certain low energy subset of the energy eigenvalues $E_{n,\ell}$
of the fully interacting system with potential $V(|\vec{r}\,|)$ in presence of
the magnetic field $B_0$ may approximately be matched onto an analogous
subset of the energy eigenvalues $E_{p,q}(n_+,n_-)$ of the
$(p,q)$-deformed quantum Landau problem, for an appropriate choice of the
parameters $p_\pm$ and $q_\pm$ as well as of the magnetic field $B$,
the electric field modulus $|\vec{E}|$ and the harmonic well curvature $k$ (in combination with
the mass and charge parameters). For example,
this should be feasible for a certain subset of the lowest energy eigenstates
of given angular momentum $\ell=n_+-n_-$, or likewise, for a certain subset
of the lowest Landau levels of both classes of systems. Prospects for such
an approximate representation of an interacting system in terms of a $(p,q)$-deformed
Landau problem in presence simply of a harmonic well which lends itself to an
exact resolution, appear to be even better than for the non-deformed case,\cite{scholtz2}
given the larger number of available parameters to be adjusted, namely
$p_\pm$ and $q_\pm$. And in all these cases, indeed the interacting potential
$V(|\vec{r}\,|)$ is traded for space noncommutativity in the fully integrable
system, namely the $(p,q)$-deformed quantum Landau problem in a spherical
harmonic well. However, space noncommutativity as derived in this Letter
is of a character more general than that associated to the ordinary quantum Hall effect
as given in (\ref{noc}). Rather, (\ref{xy}) and (\ref{eq:pxp2}) or (\ref{eq:pxp3})
define consistent extensions for space noncommutativity directly relevant to the Landau
and quantum Hall problems in their $(p,q)$-deformed quantization.

\section*{Acknowledgments}

J.~B.~G. is grateful to the Abdus Salam International Centre
for Theoretical Physics (ICTP, Trieste, Italy) for a PhD fellowship
under the grant \mbox{Prj-15}. The ICMPA is in partnership with
the Daniel Iagoniltzer Foundation (DIF), France.
J.~G. acknowledges a visiting appointment as Visiting Professor in the School of
Physics (Faculty of Science) at the University of New South Wales. He is grateful
to Prof. Chris Hamer and the School of Physics for their warm hospitality during his sabbatical
leave, and for financial support through a Fellowship of the Gordon Godfrey Fund.
His stay in Australia is also supported in part by the Belgian National
Fund for Scientific Research (F.N.R.S.) through a travel grant.
J.~G. acknowledges the Abdus Salam International Centre for Theoretical
Physics (ICTP, Trieste, Italy) Visiting Scholar Programme 
in support of a Visiting Professorship at the ICMPA. His work
is also supported by the Belgian Federal Office for Scientific, 
Technical and Cultural Affairs through the Interuniversity Attraction 
Pole (IAP) P5/27.

\end{document}